\documentclass[preprint,tightenlines,superscriptaddress,showpacs,byrevtex]{revtex4}

\usepackage{times}
\usepackage{amsmath}
\usepackage{graphicx}
\usepackage{epsfig}
\usepackage{threeparttable}
\usepackage{dcolumn}
\usepackage{bm}
\usepackage{multirow}
\usepackage{url}

\newcommand{\BR}{{\cal B}}

\newcommand{\jpsi}{J/\psi}
\newcommand{\piz}{\pi^0}

\newcommand{\EE}{e^+e^-}
\newcommand{\MM}{\mu^+\mu^-}
\newcommand{\LL}{\ell^+\ell^-}
\newcommand{\pp}{\pi^+\pi^-}

\newcommand{\beq}{\begin{equation}}
\newcommand{\eeq}{\end{equation}}
\newcommand{\bitm}{\begin{itemize}}
\newcommand{\eitm}{\end{itemize}}

\begin{document}


\title{
\boldmath Search for the $0^{--}$ Glueball in $\Upsilon(1S)$ and $\Upsilon(2S)$ decays}



\noaffiliation
\affiliation{University of the Basque Country UPV/EHU, 48080 Bilbao}
\affiliation{Beihang University, Beijing 100191}
\affiliation{Budker Institute of Nuclear Physics SB RAS, Novosibirsk 630090}
\affiliation{Faculty of Mathematics and Physics, Charles University, 121 16 Prague}
\affiliation{Chonnam National University, Kwangju 660-701}
\affiliation{University of Cincinnati, Cincinnati, Ohio 45221}
\affiliation{Deutsches Elektronen--Synchrotron, 22607 Hamburg}
\affiliation{University of Florida, Gainesville, Florida 32611}
\affiliation{Justus-Liebig-Universit\"at Gie\ss{}en, 35392 Gie\ss{}en}
\affiliation{Gifu University, Gifu 501-1193}
\affiliation{SOKENDAI (The Graduate University for Advanced Studies), Hayama 240-0193}
\affiliation{Gyeongsang National University, Chinju 660-701}
\affiliation{Hanyang University, Seoul 133-791}
\affiliation{University of Hawaii, Honolulu, Hawaii 96822}
\affiliation{High Energy Accelerator Research Organization (KEK), Tsukuba 305-0801}
\affiliation{J-PARC Branch, KEK Theory Center, High Energy Accelerator Research Organization (KEK), Tsukuba 305-0801}
\affiliation{IKERBASQUE, Basque Foundation for Science, 48013 Bilbao}
\affiliation{Indian Institute of Technology Bhubaneswar, Satya Nagar 751007}
\affiliation{Indian Institute of Technology Guwahati, Assam 781039}
\affiliation{Indian Institute of Technology Madras, Chennai 600036}
\affiliation{Institute of High Energy Physics, Chinese Academy of Sciences, Beijing 100049}
\affiliation{Institute of High Energy Physics, Vienna 1050}
\affiliation{Institute for High Energy Physics, Protvino 142281}
\affiliation{INFN - Sezione di Torino, 10125 Torino}
\affiliation{Advanced Science Research Center, Japan Atomic Energy Agency, Naka 319-1195}
\affiliation{J. Stefan Institute, 1000 Ljubljana}
\affiliation{Kanagawa University, Yokohama 221-8686}
\affiliation{Institut f\"ur Experimentelle Kernphysik, Karlsruher Institut f\"ur Technologie, 76131 Karlsruhe}
\affiliation{Kennesaw State University, Kennesaw, Georgia 30144}
\affiliation{King Abdulaziz City for Science and Technology, Riyadh 11442}
\affiliation{Department of Physics, Faculty of Science, King Abdulaziz University, Jeddah 21589}
\affiliation{Korea Institute of Science and Technology Information, Daejeon 305-806}
\affiliation{Korea University, Seoul 136-713}
\affiliation{Kyoto University, Kyoto 606-8502}
\affiliation{Kyungpook National University, Daegu 702-701}
\affiliation{\'Ecole Polytechnique F\'ed\'erale de Lausanne (EPFL), Lausanne 1015}
\affiliation{P.N. Lebedev Physical Institute of the Russian Academy of Sciences, Moscow 119991}
\affiliation{Faculty of Mathematics and Physics, University of Ljubljana, 1000 Ljubljana}
\affiliation{Ludwig Maximilians University, 80539 Munich}
\affiliation{University of Maribor, 2000 Maribor}
\affiliation{Max-Planck-Institut f\"ur Physik, 80805 M\"unchen}
\affiliation{School of Physics, University of Melbourne, Victoria 3010}
\affiliation{University of Miyazaki, Miyazaki 889-2192}
\affiliation{Moscow Physical Engineering Institute, Moscow 115409}
\affiliation{Moscow Institute of Physics and Technology, Moscow Region 141700}
\affiliation{Graduate School of Science, Nagoya University, Nagoya 464-8602}
\affiliation{Kobayashi-Maskawa Institute, Nagoya University, Nagoya 464-8602}
\affiliation{Nara Women's University, Nara 630-8506}
\affiliation{National Central University, Chung-li 32054}
\affiliation{National United University, Miao Li 36003}
\affiliation{Department of Physics, National Taiwan University, Taipei 10617}
\affiliation{H. Niewodniczanski Institute of Nuclear Physics, Krakow 31-342}
\affiliation{Nippon Dental University, Niigata 951-8580}
\affiliation{Niigata University, Niigata 950-2181}
\affiliation{Novosibirsk State University, Novosibirsk 630090}
\affiliation{Pacific Northwest National Laboratory, Richland, Washington 99352}
\affiliation{University of Pittsburgh, Pittsburgh, Pennsylvania 15260}
\affiliation{Theoretical Research Division, Nishina Center, RIKEN, Saitama 351-0198}
\affiliation{University of Science and Technology of China, Hefei 230026}
\affiliation{Showa Pharmaceutical University, Tokyo 194-8543}
\affiliation{Soongsil University, Seoul 156-743}
\affiliation{Stefan Meyer Institute for Subatomic Physics, Vienna 1090}
\affiliation{Sungkyunkwan University, Suwon 440-746}
\affiliation{School of Physics, University of Sydney, New South Wales 2006}
\affiliation{Department of Physics, Faculty of Science, University of Tabuk, Tabuk 71451}
\affiliation{Tata Institute of Fundamental Research, Mumbai 400005}
\affiliation{Excellence Cluster Universe, Technische Universit\"at M\"unchen, 85748 Garching}
\affiliation{Department of Physics, Technische Universit\"at M\"unchen, 85748 Garching}
\affiliation{Toho University, Funabashi 274-8510}
\affiliation{Department of Physics, Tohoku University, Sendai 980-8578}
\affiliation{Earthquake Research Institute, University of Tokyo, Tokyo 113-0032}
\affiliation{Department of Physics, University of Tokyo, Tokyo 113-0033}
\affiliation{Tokyo Institute of Technology, Tokyo 152-8550}
\affiliation{Tokyo Metropolitan University, Tokyo 192-0397}
\affiliation{University of Torino, 10124 Torino}
\affiliation{Virginia Polytechnic Institute and State University, Blacksburg, Virginia 24061}
\affiliation{Wayne State University, Detroit, Michigan 48202}
\affiliation{Yamagata University, Yamagata 990-8560}
\affiliation{Yonsei University, Seoul 120-749}
  \author{S.~Jia}\affiliation{Beihang University, Beijing 100191} 
  \author{C.~P.~Shen}\affiliation{Beihang University, Beijing 100191} 
  \author{C.~Z.~Yuan}\affiliation{Institute of High Energy Physics, Chinese Academy of Sciences, Beijing 100049} 
  \author{I.~Adachi}\affiliation{High Energy Accelerator Research Organization (KEK), Tsukuba 305-0801}\affiliation{SOKENDAI (The Graduate University for Advanced Studies), Hayama 240-0193} 
  \author{H.~Aihara}\affiliation{Department of Physics, University of Tokyo, Tokyo 113-0033} 
  \author{S.~Al~Said}\affiliation{Department of Physics, Faculty of Science, University of Tabuk, Tabuk 71451}\affiliation{Department of Physics, Faculty of Science, King Abdulaziz University, Jeddah 21589} 
  \author{D.~M.~Asner}\affiliation{Pacific Northwest National Laboratory, Richland, Washington 99352} 
  \author{T.~Aushev}\affiliation{Moscow Institute of Physics and Technology, Moscow Region 141700} 
  \author{R.~Ayad}\affiliation{Department of Physics, Faculty of Science, University of Tabuk, Tabuk 71451} 
  \author{V.~Babu}\affiliation{Tata Institute of Fundamental Research, Mumbai 400005} 
% ...
\collaboration{The Belle Collaboration}

\begin{abstract}

We report the first search for the $J^{PC}=0^{--}$ glueball
in $\Upsilon(1S)$ and $\Upsilon(2S)$ decays with data samples
of $(102\pm2)$ million and $(158\pm4)$ million
events, respectively, collected with the Belle detector. No significant signals
are observed in any of the proposed production modes, and the 90\%
credibility level upper limits on their branching fractions in
$\Upsilon(1S)$ and $\Upsilon(2S)$ decays are obtained. The
inclusive branching fractions of the $\Upsilon(1S)$ and
$\Upsilon(2S)$ decays into final states with a $\chi_{c1}$ are
measured to be $\BR(\Upsilon(1S)\to \chi_{c1}+ anything) =
(1.90\pm 0.43(stat.)\pm 0.14(syst.))\times 10^{-4}$ with an
improved precision over prior  measurements and
$\BR(\Upsilon(2S)\to \chi_{c1}+ anything) = (2.24\pm
0.44(stat.)\pm 0.20(syst.))\times 10^{-4}$ for the first time.

\end{abstract}


\maketitle

\section{Introduction}

The existence of bound states of gluons (so-called
``glueballs"), with a rich spectroscopy and a complex
phenomenology, is one of the early predictions of the non-abelian
nature of strong interactions described by quantum
chromodynamics (QCD)~\cite{QCD}. However, despite many years of
experimental efforts, none of these gluonic states have been established
unambiguously.
Possible reasons for this include the mixing between glueballs and conventional
mesons, the lack of solid information on the glueball production
mechanism, and the lack of knowledge about glueball
decay properties.

Of these difficulties, from the experimental point of view, the
most outstanding obstacle is the isolation of glueballs from various
quarkonium states. Fortunately, there is
a class of glueballs with three gluons and quantum numbers
incompatible with quark-antiquark bound states,
called oddballs, that are free of this conundrum. The quantum numbers of
such glueballs include $J^{PC}$ = $0^{--}$, $0^{+-}$, $1^{-+}$,
$2^{+-}$, $3^{-+}$, and so on. Among oddballs, special attention
should be paid to the $0^{--}$ state ($G_{0^{--}}$), since it is
relatively light and can be produced in the decays of vector
quarkonium or quarkoniumlike states. Two $0^{--}$ oddballs are
predicted using QCD sum rules~\cite{5} with masses of
$(3.81\pm0.12)$~GeV/$c^2$ and $(4.33\pm0.13)$~GeV/$c^2$, while the lowest-lying state
calculated using distinct bottom-up holographic models of
QCD~\cite{1507} has a mass of 2.80~GeV/$c^2$. Although the
masses have been calculated, the width and hadronic couplings to any
final states remain unknown. Possible $G_{0^{--}}$ production modes
from bottomonium decays are suggested in Ref.~\cite{5} including
$\Upsilon(1S,2S) \to \chi_{c1}+G_{0^{--}}$, $\Upsilon(1S,2S)\to
f_1(1285)+G_{0^{--}}$, $\chi_{b1} \to J/\psi+G_{0^{--}}$, and
$\chi_{b1}\to \omega+G_{0^{--}}$.

In this paper, we search for $0^{--}$ glueballs in the
production modes proposed above and define $G(2800)$, $G(3810)$,
and $G(4330)$ as the glueballs with masses
fixed at 2.800, 3.810, and
4.330~GeV/$c^2$, respectively. All the parent particles in the
above processes are copiously produced in the Belle experiment, and
may decay to the oddballs with modest rates. Since the
widths are unknown, we report an investigation of the $0^{--}$
glueballs with different assumed widths. The $\chi_{c1}$ is
reconstructed via its decays into $\gamma \jpsi$, $\jpsi\to \LL$
and $\ell=e$ or $\mu$, $f_1(1285)$ via $\eta \pp$ with $\eta\to
\gamma\gamma$, and $\omega$ via $\pp\piz$ with $\piz\to
\gamma\gamma$. As the $\chi_{c1}$ are observed clearly as tagged signals
in $\Upsilon(1S,2S)$ decays, the corresponding production
rates may be measured with improved precision.

\section{The Data Sample and Belle Detector}

This analysis utilizes the $\Upsilon(1S)$ and $\Upsilon(2S)$ data
samples with a total luminosity of $5.74$ and
$24.91~{\rm fb}^{-1}$, respectively,
corresponding to $102\times 10^6$ $\Upsilon(1S)$ and $158\times 10^6$
$\Upsilon(2S)$ events~\cite{number}.
An $89.45~{\rm fb}^{-1}$ data sample collected at $\sqrt{s} =
10.52~\mathrm{GeV}$ is used to estimate the possible irreducible
continuum contributions. Here, $\sqrt{s}$ is the center-of-mass
(C.M.) energy of the colliding $e^+e^-$ system. The data were
collected with the Belle
detector~\cite{Abashian2002117,PTEP201204D001} operated at the
KEKB asymmetric-energy
$e^{+}e^{-}$~collider~\cite{Kurokawa20031,PTEP201303A001}. Large
Monte Carlo (MC) samples of all of the investigated glueball modes
are generated with {\sc evtgen}~\cite{Lange2001152} to determine
signal line-shapes and efficiencies. The angular distribution for
$\Upsilon(2S)\to \gamma \chi_{b1}$ is simulated assuming a pure
$E1$ transition~($dN/d\cos\theta_{\gamma} \propto 1-\frac{1}{3}
\cos^2\theta_{\gamma}$~\cite{mcang}, where $\theta_\gamma$ is the
polar angle of the $\Upsilon(2S)$ radiative photon in the $\EE$
C.M. frame), and uniform phase space is used for the $\chi_{b1}$
decays. We use the uniform phase-space decay model for other decays
as well. Note that $G_{0^{--}}$ inclusive decays are modelled using
{\sc pythia}~\cite{JHEP2006.026}. Inclusive $\Upsilon(1S)$ and
$\Upsilon(2S)$ MC samples, produced using {\sc
pythia} with four times the luminosity of the
real data, are used to identify possible peaking backgrounds from
$\Upsilon(1S)$ and $\Upsilon(2S)$ decays.

The Belle detector is a large solid-angle magnetic spectrometer
that consists of a silicon vertex detector, a 50-layer central
drift chamber (CDC), an array of aerogel threshold Cherenkov
counters (ACC), a barrel-like arrangement of time-of-flight
scintillation counters (TOF), and an electromagnetic calorimeter
comprised of CsI(Tl) crystals (ECL) located inside a
superconducting solenoid coil that provides a $1.5~\hbox{T}$
magnetic field. An iron flux-return yoke located outside the coil
is used to detect $K^{0}_{L}$ mesons and to identify
muons. A detailed description of the Belle detector can be found
in Refs.~\cite{Abashian2002117, PTEP201204D001}.

\section{Event Selection}

Charged tracks from the primary vertex with $dr<0.5~\mathrm{cm}$
and $|dz|<4~\mathrm{cm}$ are selected, where $dr$ and $dz$ are the
impact parameters perpendicular to and along the beam direction,
respectively, with respect to the interaction point. In addition,
the transverse momentum of every charged track in the laboratory
frame is restricted to be larger than $0.1~\mathrm{GeV/{\mathit
c}}$. We require the number of well-reconstructed charged tracks
to be greater than four to suppress the significant background
from quantum electrodynamics  processes. For charged tracks, information from different
detector subsystems including specific ionization in the CDC, time
measurements in the TOF and the response of the ACC is combined to
form the likelihood ${\mathcal L}_i$ for particle species $i$,
where $i=\pi$,~$K$, or $p$~\cite{like}. Charged tracks with
$R_{K}=\mathcal{L}_{K}/(\mathcal{L}_K+\mathcal{L}_\pi)<0.4$ are
considered to be pions. With this condition, the pion identification
efficiency is $96\%$ and the kaon misidentification
rate is about $9\%$. A similar likelihood ratio is defined as
$R_e=\mathcal{L}_e/(\mathcal{L}_e+\mathcal{L}_{{\rm
non}-e})$~\cite{Hanagaki485490} for electron identification and
$R_{\mu}=\mathcal{L}_{\mu}/(\mathcal{L}_{\mu}+\mathcal{L}_{K}
+\mathcal{L}_{\pi})$~\cite{Abashian49169} for muon identification.
An ECL cluster is taken as a photon candidate if it does not
match the extrapolation of any charged track and its energy is
greater than 50~MeV.

\begin{figure*}[htbp]
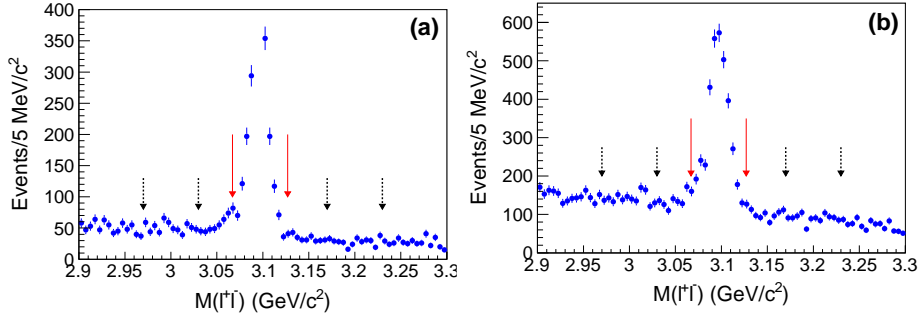

\includegraphics[height=6cm,angle=-90]{fig1a.epsi}
\includegraphics[height=6cm,angle=-90]{fig1b.epsi}
\caption{The $\ell^+ \ell^-$ invariant mass distributions in the
$\Upsilon(1S)$ (a) and $\Upsilon(2S)$ (b) data samples. The solid arrows
show the $\jpsi$ signal region, and the dashed arrows show the $\jpsi$ mass sideband regions.}\label{Mll-Y12S}
\end{figure*}

To reduce the effect of bremsstrahlung and final-state radiation,
photons detected in the ECL within a $50~\mathrm{mrad}$ cone of
the original electron or positron direction are included in the
calculation of the $e^+/e^-$ four-momentum. For the lepton pair
$\LL$ used to reconstruct the $\jpsi$, both of the tracks should have
$R_e>0.95$ in the $\EE$ mode; or one track should have
$R_\mu>0.95$ and the other $R_\mu>0.05$ in the $\MM$ mode. The
lepton pair identification efficiencies for $e^+e^-$ and
$\mu^+\mu^-$ are $96\%$ and $93\%$, respectively. After
all  event selection requirements, significant $J/\psi$ signals
are seen in the $\Upsilon(1S)$ and $\Upsilon(2S)$ data samples, as
shown in Figs.~\ref{Mll-Y12S} (a) and (b). Since different modes
have almost the same $\jpsi$ mass resolutions, we define the
$\jpsi$ signal region in the window
$|M_{\ell^+\ell^-}-m_{\jpsi}|<0.03$~GeV/$c^2$ ($\sim 2.5\sigma$)
indicated by the arrows, where $m_{\jpsi}$ is the $\jpsi$ nominal
mass~\cite{jpsimass}, while the $\jpsi$ mass sideband is
$2.97~\hbox{GeV}/c^2<M_{\ell^+\ell^-}<3.03$~GeV/$c^2$  or
$3.17~\hbox{GeV}/c^2<M_{\ell^+\ell^-}<3.23$~GeV/$c^2$, which is twice as wide as
the signal region. In order to
improve the $\jpsi$ momentum resolution,
a mass-constrained fit is applied to the
$\jpsi$ candidates in the signal region.

\section{MEASUREMENTS OF $\Upsilon(1S,2S\bf) \to \chi_{c1}+ anything$}

\begin{figure*}[htbp]
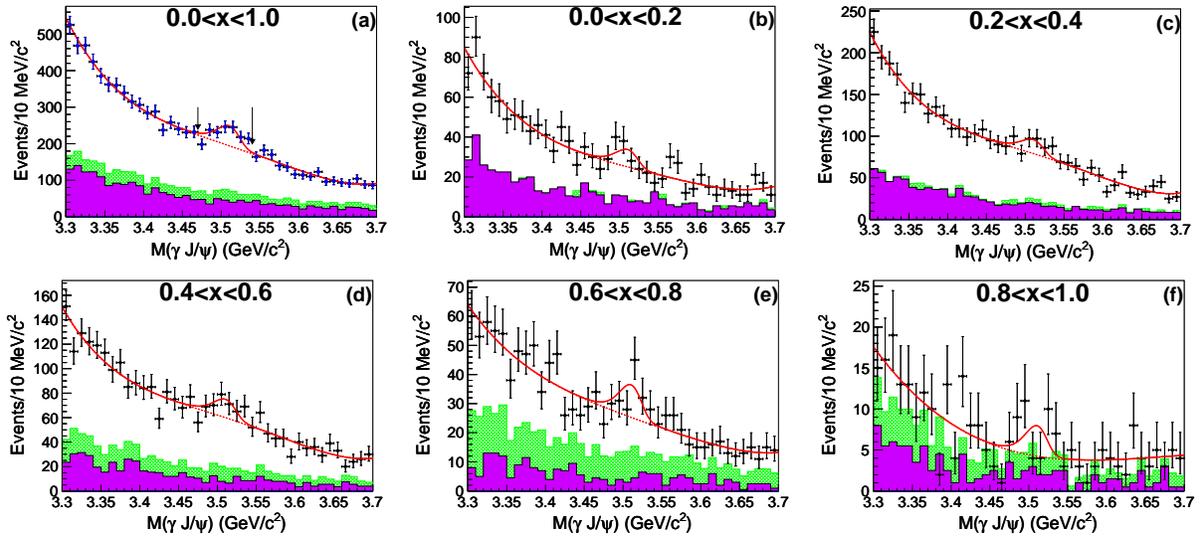

\includegraphics[height=5cm,angle=-90]{fig2a.epsi}
\hspace{0.2cm}
\includegraphics[height=5cm,angle=-90]{fig2b.epsi}
\hspace{0.2cm}
\includegraphics[height=5cm,angle=-90]{fig2c.epsi}
\\\vspace{0.2cm}
\includegraphics[height=5cm,angle=-90]{fig2d.epsi}
\hspace{0.2cm}
\includegraphics[height=5cm,angle=-90]{fig2e.epsi}
\hspace{0.2cm}
\includegraphics[height=5cm,angle=-90]{fig2f.epsi}
\caption{(Color online) Invariant mass distributions of the $\chi_{c1}$
candidates in the entire $x$ region (a) and for $x$ bins of size
$0.2$ (b--f). The dots with error bars are the $\Upsilon(1S)$
data. The solid lines are the best fits, and the dotted lines represent the backgrounds. The shaded histograms are from
the normalized $\jpsi$
mass sidebands and cross-hatched histograms are from the normalized
continuum contributions described in the text.
The arrows in (a) show the $\chi_{c1}$ signal region that will be used to search for glueballs in the channel
$\Upsilon(1S) \to \chi_{c1}+G_{0^{--}}$ below.}
\label{12Sdata1}
\end{figure*}

Before searching for the $G_{0^{--}}$ in $\Upsilon(1S,2S) \to
\chi_{c1}+G_{0^{--}}$, we measure the inclusive $\chi_{c1}$
production in $\Upsilon(1S,2S)$. The $\jpsi$ candidate is combined
with any one of the photon candidates to reconstruct the $\chi_{c1}$
signal. The $\gamma \jpsi$ invariant mass distributions for the
$\chi_{c1}$ candidates are shown in Figs.~\ref{12Sdata1}(a) and
~\ref{12Sdata2}(a) from $\Upsilon(1S)$ and $\Upsilon(2S)$ decays,
respectively. Clear $\chi_{c1}$ signals are observed in both data
samples, while no clear $\chi_{c2}$ signals are seen.
No evidence for $\chi_{c1}$ signals is found in the $\jpsi$-mass
sideband events nor the continuum data sample, as can be seen from
the same plots.

\begin{figure*}[htbp]
\includegraphics[height=5cm,angle=-90]{fig3a.epsi}
\hspace{0.2cm}
\includegraphics[height=5cm,angle=-90]{fig3b.epsi}
\hspace{0.2cm}
\includegraphics[height=5cm,angle=-90]{fig3c.epsi}
\\\vspace{0.2cm}
\includegraphics[height=5cm,angle=-90]{fig3d.epsi}
\hspace{0.2cm}
\includegraphics[height=5cm,angle=-90]{fig3e.epsi}
\hspace{0.2cm}
\includegraphics[height=5cm,angle=-90]{fig3f.epsi}
\caption{(Color online) Invariant mass distributions of the $\chi_{c1}$
candidates in the entire $x$ region (a) and for $x$ bins of size
$0.2$ (b--f). The dots with error bars are the $\Upsilon(2S)$
data.  The solid lines are
the best fits, and the dotted lines represent the backgrounds.
The shaded histograms are from the normalized $\jpsi$
mass sidebands and cross-hatched histograms are from the normalized
continuum contributions described in the text.
The arrows in (a) show the $\chi_{c1}$ signal region
that will be used to search for glueballs in the channel $\Upsilon(2S) \to \chi_{c1}+G_{0^{--}}$ below.}
\label{12Sdata2}
\end{figure*}

The continuum background contribution is determined using a large
amount of data taken at $\sqrt{s}=10.52$~GeV, extrapolated
down to the lower resonances. The scale factor used for this
extrapolation is $f_{{\rm scale}} =
\mathcal{L}_{\Upsilon}/\mathcal{L}_{\rm con}\times
\sigma_{\Upsilon}/\sigma_{\rm con}\times
\varepsilon_{\Upsilon}/\varepsilon_{\rm con}$, where
$\mathcal{L}_{\Upsilon}/\mathcal{L}_{\rm con}$,
$\sigma_{\Upsilon}/\sigma_{\rm con}$, and
$\varepsilon_{\Upsilon}/\varepsilon_{\rm con}$ are the ratios of
the integrated luminosities, cross sections, and efficiencies,
respectively, for the $\Upsilon$ and continuum samples. The cross section
extrapolation with beam energy is assumed to have a
$1/s^{2}$~\cite{PhysRevD.69.094027,PhysRevD.56.321,hep-hp13108597}
dependence. Contributions from $e^+e^-$
annihilation without $\jpsi$ events have been subtracted to
avoid double counting of continuum events. The resulting scale
factor is about 0.10 for $\Upsilon(1S)$ and 0.35 for
$\Upsilon(2S)$ decays. For $\Upsilon(2S)\to \chi_{c1}+anything$,
another background is the intermediate transition $\Upsilon(2S)
\to \pi^+ \pi^- \Upsilon(1S)$ or $\pi^0 \pi^0 \Upsilon(1S)$ with
$\Upsilon(1S)$ decaying into $\chi_{c1}$.
Such contamination is
removed by requiring the $\pi \pi$ recoil mass
to be outside the $[9.45, 9.47]$~GeV/$c^2$ region
for all $\pi\pi$ combinations.

Considering the slight differences in the MC-determined
reconstruction efficiencies for different $\chi_{c1}$ momenta, we
partition the data samples according to the scaled momentum
$x=p^{\ast}_{\chi_{c1}}/(\frac{1}{2\sqrt{s}}\times(s-m_{\chi_{c1}}^2))$~\cite{PhysRevD.70.072001},
where $p^{\ast}_{\chi_{c1}}$ is the momentum of the $\chi_{c1}$
candidate in the $e^+e^-$ C.M. system, and $m_{\chi_{c1}}$ is the
$\chi_{c1}$ nominal mass~\cite{PDG}. The value of
$\frac{1}{2\sqrt{s}}\times (s-m_{\chi_{c1}}^2)$ is the value of
$p^{\ast}_{\chi_{c1}}$ for the case where the $\chi_{c1}$
candidate recoils against a massless particle. The use of $x$
removes the beam-energy dependence in comparing the continuum data
to that taken at the $\Upsilon(1S,2S)$ resonances. The $\gamma
\jpsi$ invariant mass distribution in each $\Delta x = 0.2$ bin
is shown in Figs.~\ref{12Sdata1}(b--f) and ~\ref{12Sdata2}(b--f)
for $\Upsilon(1S)$ and $\Upsilon(2S)$ decays, respectively.

\begin{figure}[htpb]
\includegraphics[height=6cm, angle=-90]{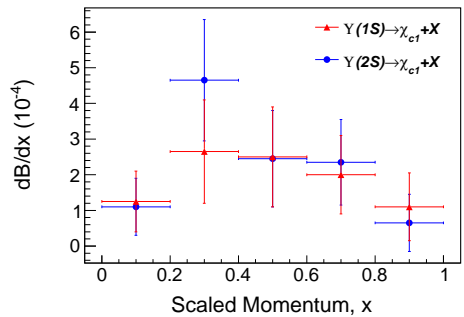}
\caption{Differential branching fractions for $\Upsilon(1S)$ and
$\Upsilon(2S)$ inclusive decays into $\chi_{c1}$ as a function of the
scaled momentum $x$, defined in the text.
The error bar of each point
is the sum in quadratic of the statistical and systematic errors.
  }\label{11111}
\end{figure}

\begin{figure*}[htbp]
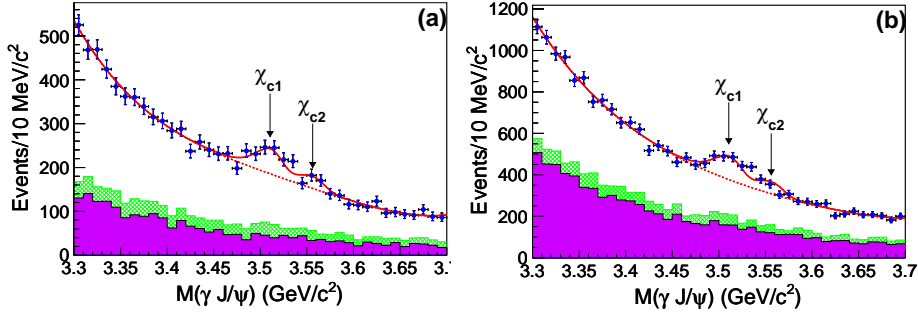

\includegraphics[height=6cm,angle=-90]{fig5a.epsi}
\includegraphics[height=6cm,angle=-90]{fig5b.epsi}
\caption{(Color online)
The $\gamma J/\psi$ invariant mass distributions in the entire $x$ region in $\Upsilon(1S)$ (a)
and $\Upsilon(2S)$ (b) data. The dots with error bars are the $\Upsilon(1S,2S)$ data. The arrows show
the expected positions of the $\chi_{c1}$ and $\chi_{c2}$ signals. The solid lines are the best fits with
the $\chi_{c1}$ and $\chi_{c2}$ signals included, and the dotted lines represent the backgrounds. The shaded
histograms are from the normalized $\jpsi$ mass sidebands and cross-hatched histograms are from the
normalized continuum contributions described in the text.}\label{12schic2}
\end{figure*}

An unbinned extended likelihood fit is applied to the
$x$-dependent $\chi_{c1}$ spectra to extract the signal yields in
the $\Upsilon(1S)$ or $\Upsilon(2S)$ data sample.
Due to the slight dependence on momentum, the $\chi_{c1}$ shape in each
$x$ bin is described by a Breit-Wigner (BW) function convolved with a Novosibirsk function~\cite{Nov},
where all parameter values are fixed to those from the fit to the MC-simulated $\chi_{c1}$ signal.
Since no peaking backgrounds are found, a third-order Chebyshev polynomial shape is used for the backgrounds.
The fit results are shown
in Figs.~\ref{12Sdata1} and \ref{12Sdata2}, and the fitted
$\chi_{c1}$ signal yields ($N_{\rm fit}$) in the entire $x$ region
and each $x$ bin from $\Upsilon(1S)$ and $\Upsilon(2S)$ decays are
itemized in Table~\ref{Table-Nfit}, together with the
reconstruction efficiencies from MC signal simulations ($\varepsilon$), the total systematic
uncertainties ($\sigma_{\rm syst}$)---which
are the sum of the common systematic errors (discussed
below)---and fit errors estimated in each $x$ bin or
the full range in $x$, and the
corresponding branching fractions ($\BR$). The total numbers of
$\chi_{c1}$ events, \textit{i.e.}, the sums of the signal yields in all of
the $x$ bins, the sums of the
$x$-dependent efficiencies weighted by the signal fraction in that
$x$ bin, and the measured branching
fractions are listed in the bottom row. In comparison with the
previous result of $(2.3\pm 0.7)\times
10^{-4}$~\cite{PhysRevD.70.072001} for $\Upsilon(1S)\to \chi_{c1}+
anything$, our measurement of $(1.90\pm 0.43(stat.)\pm 0.14(syst.))\times 10^{-4}$ has an improved precision
and lower continuum background due to
the requirement that the number of charged tracks be greater than four.
The branching fraction for $\Upsilon(2S)\to \chi_{c1}+
anything$ is measured for the first time and found to be
$(2.24\pm 0.44(stat.)\pm 0.20(syst.))\times 10^{-4}$.
The differential branching fractions of $\Upsilon(1S,2S)$ decays
into $\chi_{c1}$ are shown in Fig.~\ref{11111}.
A fit with an additional $\chi_{c2}$ signal shape is also performed
in the entire $x$ region in
the $\Upsilon(1S)$ or $\Upsilon(2S)$ data sample,
as shown in Fig.~\ref{12schic2}.
The difference in the number of fitted $\chi_{c1}$
yields is included in the systematic error.
The  $\chi_{c2}$ signal significance from the fit
 is less than $2.7\sigma$ ($3.2\sigma$)
in the $\Upsilon(1S)$ ($\Upsilon(2S)$) data sample.
The 90\% credibility level (C.L.)~\cite{cl} upper limit (measured as described below)
for the $\Upsilon(1S)\to \chi_{c2}+ anything$ branching fraction is $3.09\times10^{-4}$, with systematic errors included, to be
compared with the previous result of $(3.4\pm1.0)\times 10^{-4}$~\cite{PhysRevD.70.072001},
and the measured $\Upsilon(2S)\to \chi_{c2}+ anything$ branching fraction
is $(2.28\pm 0.73(stat.)\pm 0.34(syst.))\times 10^{-4}$ ($<3.28\times 10^{-4}$ at 90\% C.L.).

\begin{table*}[t]
\begin{threeparttable}
\caption{\label{Table-Nfit} Summary of the branching fraction
measurements of $\Upsilon(1S,2S)$ inclusive decays into
$\chi_{c1}$, where $N_{\rm fit}$ is the number of fitted signal
events, $\varepsilon$ (\%) is the reconstruction efficiency,
$\sigma_{syst}$ (\%) is the total systematic error on the
branching fraction measurement, and $\BR$ is the measured
branching fraction.}
\footnotesize
  \begin{tabular}{c|r@{$\pm$}lccc||r@{$\pm$}lccc}
  \hline\hline
    \multicolumn{6}{c}{$\Upsilon(1S)\to \chi_{c1}+{\rm anything}$} & \multicolumn{5}{c}{$\Upsilon(2S)\to\chi_{c1}+{\rm anything}$} \\[-2pt]
   $x$ & \multicolumn{2}{c}{$N_{\rm fit}$} & $\varepsilon (\%)$ & $\sigma_{\rm syst} (\%)$ & $\mathcal{B}(10^{-4})$& \multicolumn{2}{c}{$N_{\rm fit}$} & $\varepsilon (\%)$
   & $\sigma_{\rm syst}(\%)$ & $\mathcal{B}(10^{-4})$ \\
  \hline
    $(0.0,0.2)$  &$34.0$  & $18.0$ & $31.77$ & $17.0$ &  $0.25\pm0.13\pm0.04$ & $43.0$  & $25.1$ & $30.56$ & $15.6$ &  $0.22\pm0.13\pm0.03$ \\
    $(0.2,0.4)$  &$65.2$ & $30.7$ & $29.09$ & $7.2$  &  $0.53\pm0.25\pm0.04$ & $161.3$  & $44.1$ & $27.11$ & $9.6$  &  $0.93\pm0.25\pm0.09$ \\
    $(0.4,0.6)$  &$58.4$  & $26.9$ & $27.70$ & $9.5$  &  $0.50\pm0.23\pm0.05$ & $85.5$  & $39.0$ & $26.50$ & $9.6$  &  $0.49\pm0.22\pm0.05$ \\
    $(0.6,0.8)$  & $43.4$  & $18.3$ & $25.72$  & $13.0$ & $0.40\pm0.17\pm0.05$ & $72.7$  & $28.5$ & $24.25$  & $12.6$ & $0.47\pm0.18\pm0.06$ \\
    $(0.8,1.0)$  &$14.4$   & $9.5$ & $15.35$  & $22.3$  & $0.22\pm0.15\pm0.05$ & $13.1$  & $14.2$  & $15.69$  & $17.4$  & $0.13\pm0.14\pm0.02$ \\
    All $x$  &$215.4$ & $49.2$ & $27.54$  &  $7.1$  & $1.90\pm0.43\pm0.14$ & $375.6$  & $73.2$ & $26.41$  &  $9.1$  & $2.24\pm0.44\pm0.20$ \\
  \hline\hline
  \end{tabular}
  \end{threeparttable}
\end{table*}

\section{Search for $0^{--}$ Glueballs in $\Upsilon(1S)$, $\Upsilon(2S)$,
and $\chi_{b1}$ decays}

In the channels $\Upsilon(1S,2S) \to
\chi_{c1}+G_{0^{--}}$, $\Upsilon(1S,2S)\to f_1(1285)+G_{0^{--}}$,
$\chi_{b1} \to J/\psi+G_{0^{--}}$, and $\chi_{b1}\to
\omega+G_{0^{--}}$, we search for the $G_{0^{--}}$ signals in
the recoil mass spectra of the $\chi_{c1}$, $f_1(1285)$, $\jpsi$, and
$\omega$ with $G_{0^{--}}$ widths varying from 0.0 to 0.5~GeV in
steps of 0.05~GeV.
After all selection requirements, no peaking backgrounds are found in the $\chi_{c1}$,
$f_1(1285)$, $\jpsi$, or $\omega$ mass sideband events, or in the
continuum production in the $G_{0^{--}}$ signal regions, in
agreement with the expectation according to the $\Upsilon(1S,2S)$
generic MC samples.

An unbinned extended maximum-likelihood fit to all the recoil mass
spectra is performed to extract the signal and background yields
in the $\Upsilon(1S)$ and $\Upsilon(2S)$ data samples.
The signal shapes of the $G_{0^{--}}$ signals used in the fits are obtained
directly from MC simulations, while for the background a third-order
Chebyshev polynomial function is adopted.
In each fit, only one glueball candidate with
fixed mass and width is included and the upper limit on the number of signal events is obtained.


\subsection{MEASUREMENTS OF $\Upsilon(1S,2S) \to \chi_{c1}+G_{0^{--}}$}

\begin{figure*}[htbp]
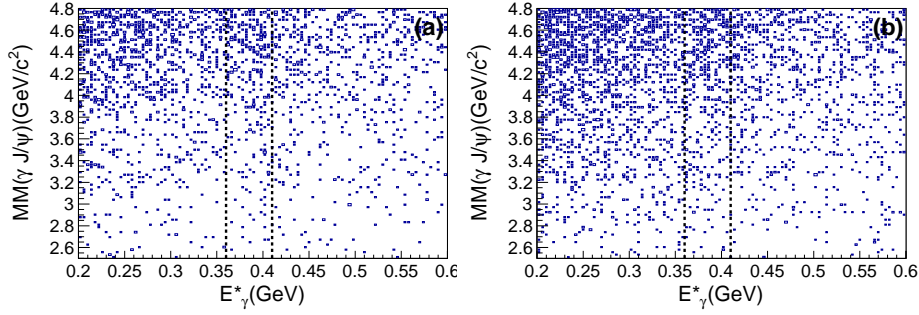

\includegraphics[height=6cm,angle=-90]{fig6a.epsi}
\includegraphics[height=6cm,angle=-90]{fig6b.epsi}
\caption{Scatter plots of the recoil
mass of $\gamma \jpsi$ versus the photon energy from $\chi_{c1}$
radiative decays in the $\gamma \jpsi$ C.M. frame in $\Upsilon(1S)$ (a) and $\Upsilon(2S)$
(b) data. The dotted lines show the expected $\chi_{c1}$ signal
region.}\label{jj-scatter}
\end{figure*}

\begin{figure*}[htbp]
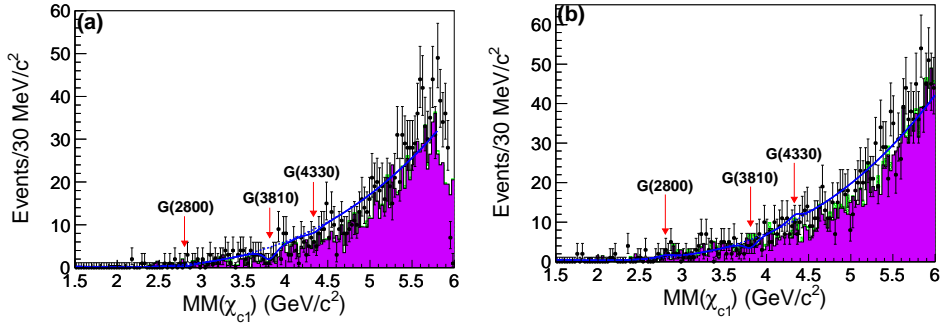

\includegraphics[height=6cm,angle=-90]{fig7a.epsi}
\hspace{0.2cm}
\includegraphics[height=6cm,angle=-90]{fig7b.epsi}
\caption{(Color online)
The $\chi_{c1}$ recoil mass spectra in the $\Upsilon(1S)$ (a) and $\Upsilon(2S)$ (b) data samples.
The solid curves show the results of the fit described in the text, including the
$G(2800)$, $G(3810)$, and $G(4330)$ states, with a common width fixed to 0.15 GeV (for illustration) and
with central values indicated by the arrows. The dashed curves show the fitted background.
The shaded histograms are from the normalized $\chi_{c1}$ mass sideband events and the cross-hatched
histograms show the normalized continuum contributions.}
\label{Y12Sdata}
\end{figure*}

For $\Upsilon(1S,2S) \to \chi_{c1}+G_{0^{--}}$,
Figs.~\ref{jj-scatter}(a) and (b) show the scatter plots of
the $\gamma \jpsi$ recoil mass versus
the energy of the photon in the $\gamma
\jpsi$ C.M. frame in the
$\Upsilon(1S)$ and $\Upsilon(2S)$ data samples, respectively. We require
the photon energy from $\chi_{c1}$ radiative decays in the $\gamma
\jpsi$ C.M. frame to satisfy $0.36~\hbox{GeV}<E_{\gamma}^{*}<0.41$~GeV to suppress
the non-$\chi_{c1}$ backgrounds. The $\chi_{c1}$ mass
sidebands are defined as $0.25~\hbox{GeV}<E_{\gamma}^{*}<0.28~\hbox{GeV}$ or
$0.43~\hbox{GeV}<E_{\gamma}^{*}<0.50$~GeV. After the application of the above
requirements, Fig.~\ref{Y12Sdata} shows the recoil mass spectra
of $\chi_{c1}$ candidates in the $\Upsilon(1S,2S)$ data. There are no
evident signals for any of the $G_{0^{--}}$ states at any of the expected
positions.
Since the width is unknown, the fit is repeated with $G_{0^{--}}$ widths from 0 to
0.5 GeV in steps of 0.05 GeV.
The fit results for the $G(2800)$, $G(3810)$, and $G(4330)$
signals with their widths fixed at 0.15~GeV are shown in
Fig.~\ref{Y12Sdata} as an example.
The fit yields $-3.8\pm3.9$ ($6.2\pm 6.4$), $-20.4\pm 7.8$ ($-18.5\pm9.2$),
and $-5.7\pm 11.3$ ($12.5\pm14.9$) events for the $G(2800)$, $G(3810)$, and
$G(4330)$ signals, respectively, in the $\Upsilon(1S)$ ($\Upsilon(2S)$) data sample.

Since the statistical significance in each case is less than
$3\sigma$, upper limits on the numbers of signal events, $N^{\rm
UL}$, are determined at the $90\%$ C.L. by
solving the equation $\int^{N^{\rm UL}}_0 \mathcal{L}(x)dx /
\int^{+\infty}_0\mathcal{L}(x)dx = 0.9$, where $x$ is
the number of signal events and $\mathcal{L}(x)$
is the maximized likelihood of the data assuming $x$ signal events.
The signal significances
are calculated using $\sqrt{-2\ln(\mathcal{L}(0)/\mathcal{L}_{\rm
max})}$, where $\mathcal{L}_{\rm max}$ is the maximum of $\mathcal{L}(x)$.
To take into account systematic uncertainties
discussed below, the above likelihood is convolved with a
Gaussian function whose width equals the total systematic
uncertainty.

\begin{figure*}[htbp]
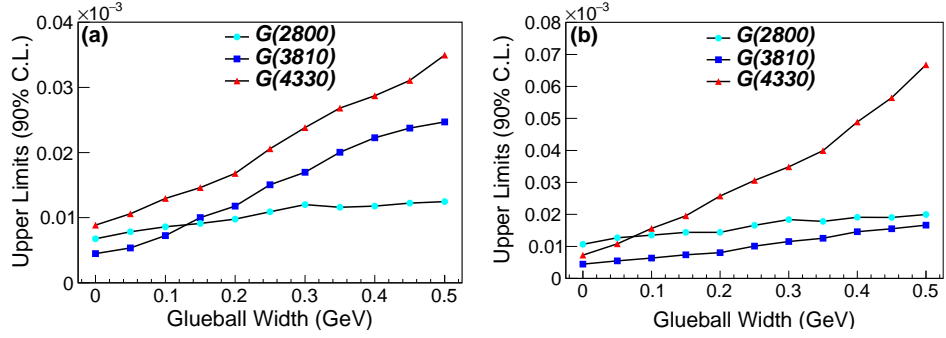

\includegraphics[height=6cm,angle=-90]{fig8a.epsi}
\hspace{0.3cm}
\includegraphics[height=6cm,angle=-90]{fig8b.epsi}
\caption{ (Color online) The upper limits on the branching fractions for
$\Upsilon(1S)\to \chi_{c1}+G_{0^{--}}$ (a) and $\Upsilon(2S)\to
\chi_{c1}+G_{0^{--}}$ (b)
as a function of the assumed $G_{0^{--}}$ decay width.}\label{final1}
\end{figure*}

The calculated upper limits on the numbers of signal events
($N^{\rm UL}$) and branching fraction ($\BR^{\rm UL}$) with widths from 0.0 to 0.5~GeV
for each $G_{0^{--}}$ state are listed in
Table~\ref{Table-X-Summary}, together with the reconstruction
efficiencies ($\varepsilon$) and the systematic
uncertainties ($\sigma_{\rm syst}$).
The results are displayed graphically in Fig.~\ref{final1}.


\subsection{MEASUREMENTS OF $\Upsilon(1S,2S) \to f_1(1285)+G_{0^{--}}$}

\begin{figure*}[htbp]
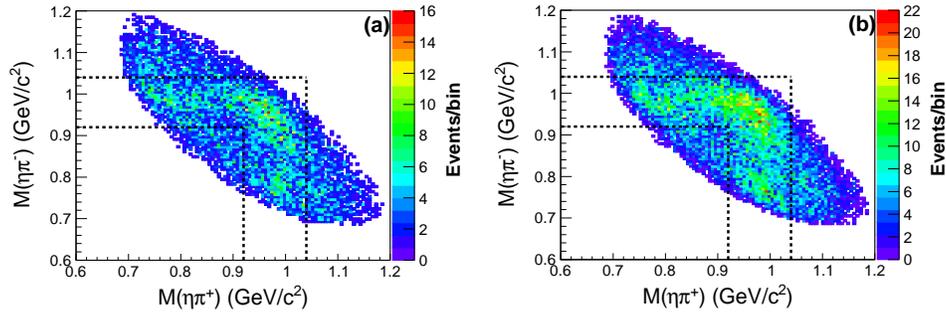

\includegraphics[height=6cm,angle=-90]{fig9a.epsi}\hspace{0.35cm}
\includegraphics[height=6cm,angle=-90]{fig9b.epsi}
\caption{(Color online) Scatter plots of $M(\eta\pi^-)$ versus $M(\eta\pi^+)$ in
$\Upsilon(1S)$ (a) and $\Upsilon(2S)$ (b) data. The dotted lines
show the $a_0(980)$ signal region.}\label{Dalitz}
\end{figure*}

\begin{figure*}[htbp]
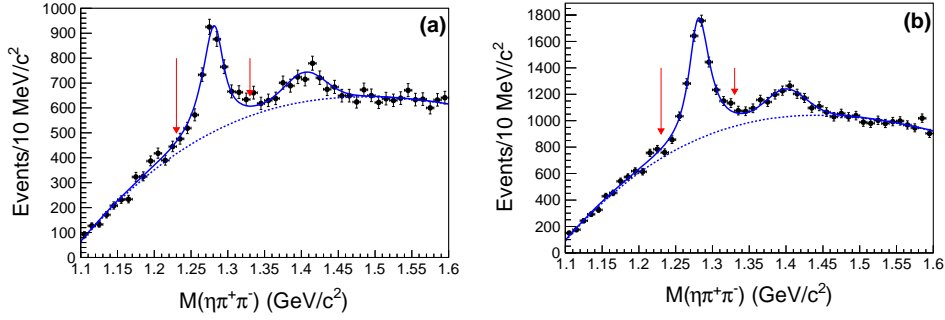

\includegraphics[height=6cm,angle=-90]{fig10a.epsi}\hspace{0.35cm}
\includegraphics[height=6cm,angle=-90]{fig10b.epsi}
\caption{The $\eta\pi^+\pi^-$ invariant mass spectra in
$\Upsilon(1S)$ (a) and $\Upsilon(2S)$ (b) data with the $\eta \pi$
mass within the $a_0(980)$ mass region.
The solid lines are the best fits and the dotted lines represent
the backgrounds.
The red arrows show the
$f_1(1285)$ signal region.}\label{f1_f1285}
\end{figure*}

Candidate $f_1(1285)$ states are reconstructed via $\eta\pi^+\pi^-$, $\eta\to \gamma \gamma$.
The energies of the photons from the $\eta$ decays are required to be
greater than 0.25~GeV to suppress background photons.
The photons from possible $\pi^0$ decays
are vetoed if the invariant mass of one photon from the $\eta$ candidate and any other photon
satisfies $|M(\gamma\gamma)-m_{\pi^0}|<18$~MeV/$c^2$, where
$m_{\pi^0}$ is the $\pi^0$ nominal mass. We perform a mass-constrained
kinematic fit to the surviving $\eta$ candidates and require $\chi^{2}<10$. A
clear $K^{0}_{S}$ signal is seen in the $\pi^+\pi^-$ invariant
mass distribution and such backgrounds are removed
by requiring that the $\pi^+\pi^-$ mass not fall
between 0.475 and 0.515~GeV/$c^2$.
After the application of these requirements, the
scatter plots of the $\eta\pi^-$ invariant mass versus the
$\eta\pi^+$ invariant mass in $\Upsilon(1S)$ and $\Upsilon(2S)$ data are shown in
Figs.~\ref{Dalitz}(a) and (b), respectively; here, $a_0(980)$
signals are observed. Since the $f_1(1285)$ decays into $\eta\pp$
primarily via the $a_0(980)\pi$ intermediate state, we require either
$M(\eta\pi^+)$ or $M(\eta\pi^-)$ to be in a $\pm$60~MeV/$c^2$ mass
window centered on the $a_0(980)$ nominal mass. The
$\eta\pi^+\pi^-$ invariant mass spectra are shown in
Fig.~\ref{f1_f1285}; clear $f_1(1285)$ and $\eta(1405)$ signals
are observed. BW functions are convolved with Novosibirsk functions for the
$f_1(1285)$ and $\eta(1405)$ signal shapes and a third-order Chebychev
function is taken for the background shape in the fits to the
$\eta\pi^+\pi^-$ invariant mass spectra. The fit results are shown
in Fig.~\ref{f1_f1285} as the solid lines. We define the
$f_1(1285)$ signal region as
$1.23~\hbox{GeV}/c^2<M(\eta\pi^+\pi^-)<1.33$~GeV/$c^2$ and its mass sideband as
$1.50~\hbox{GeV}/c^2<M(\eta\pi^+\pi^-)<1.60$~GeV/$c^2$.

\begin{figure*}[htbp]
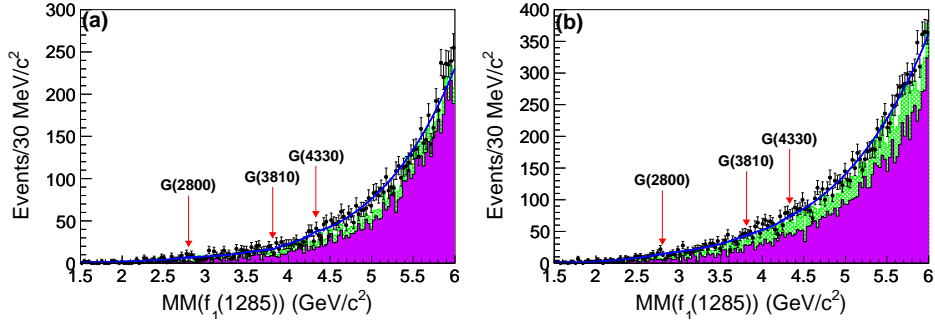

\includegraphics[height=6cm,angle=-90]{fig11a.epsi}
\hspace{0.2cm}
\includegraphics[height=6cm,angle=-90]{fig11b.epsi}
\caption{(Color online)
The $f_1(1285)$ recoil mass spectra in the $\Upsilon(1S)$ (a) and $\Upsilon(2S)$ (b) data samples.
The solid curves show the results of the fit described in the text, including the
$G(2800)$, $G(3810)$, and $G(4330)$ states, with a common width fixed to 0.15 GeV and
with central values indicated by the arrows. The dashed curves show the fitted background.
The shaded histograms are from the normalized $f_1(1285)$ mass sideband events and the cross-hatched
histograms show the normalized continuum contributions.}
\label{f1data}
\end{figure*}

After applying all of the above requirements, Fig.~\ref{f1data} shows
the recoil mass spectra of the $f_1(1285)$ in $\Upsilon(1S,2S)$ data,
together with the background from the normalized $f_1(1285)$ mass sideband
events and the normalized continuum contributions. No
evident $G_{0^{--}}$ signals are seen.
An unbinned extended maximum-likelihood fit, repeated with $G_{0^{--}}$ widths from 0 to
0.5 GeV in steps of 0.05 GeV, is applied to the recoil mass spectra.
The results of illustrative fits including $G(2800)$, $G(3810)$, and $G(4330)$ signals
with widths fixed at 0.15~GeV are shown in Fig.~\ref{f1data}.
The fits yield $20.2\pm 14.2$ ($25.0\pm 22.3$) $G(2800)$ signal
events, $-23.0\pm 25.2$ ($31.7\pm 39.0$) $G(3810)$ signal events,
and $31.8\pm 30.0$ ($68.3\pm 47.2$) $G(4330)$ signal events in
$\Upsilon(1S)$ ($\Upsilon(2S)$) data.

\subsection{MEASUREMENTS OF $\chi_{b1} \to \jpsi+G_{0^{--}}$}

The  $\chi_{b1}$ is identified through the decay $\Upsilon(2S)\to \gamma \chi_{b1}$.
Figure~\ref{jj-scatter2} shows
the scatter plot of the recoil mass of $\gamma
\jpsi$ versus
the energy of the $\Upsilon(2S)$ radiative
photon in the $\EE$ C.M. frame and the $E_{\gamma}^{*}$ distribution. To select the $\chi_{b1}$ signal, we require
$0.115~\hbox{GeV}<E_{\gamma}^{*}<0.145$~GeV. Figure~\ref{GBdata} shows the
recoil mass spectrum of $\gamma \jpsi$ in $\Upsilon(2S)$ data after
all of the above selections, together with the background estimated
from the normalized $\jpsi$ mass sideband events and the
normalized continuum contributions. No evident $G_{0^{--}}$
signal is observed. An unbinned extended maximum-likelihood fit
is applied to the $\gamma \jpsi$ recoil mass spectrum.
The result of a typical fit including $G(2800)$, $G(3810)$, and $G(4330)$ signals with
widths fixed at 0.15~GeV is shown in Fig.~\ref{GBdata}. The fit
yields $-11.4\pm 6.8$ $G(2800)$ signal events, $-7.1\pm 13.5$
$G(3810)$ signal events, and $27.0\pm 19.5$ $G(4330)$ signal
events.

\begin{figure}[htpb]
\includegraphics[height=6cm,angle=-90]{fig12a.epsi}
\hspace{0.35cm}
\includegraphics[height=6cm,angle=-90]{fig12b.epsi}
\caption{Scatter plot of the recoil mass of
$\gamma \jpsi$ versus the energy of the $\Upsilon(2S)$
radiative photon in the $\EE$ C.M. frame (a) and the distribution of the $\Upsilon(2S)$ radiative
photon's energy (b). The dotted lines indicate the expected $\chi_{b1}$
signal region. The arrow shows the position of $\chi_{b1}$.}\label{jj-scatter2}
\end{figure}

\begin{figure*}[htbp]
\includegraphics[height=7cm,angle=-90]{fig13.epsi}
\caption{(Color online)
The $\gamma\jpsi$ recoil mass spectrum
for $\Upsilon(2S)\to \gamma \chi_{b1} \to \gamma \jpsi + anything$
in the $\Upsilon(2S)$ data sample.
The solid curve shows the result of the fit described in the text, including the
$G(2800)$, $G(3810)$, and $G(4330)$ states, with a common width fixed to 0.15 GeV and
with central values indicated by the arrows. The dashed curve shows the fitted background.
The shaded histogram is from the normalized $\jpsi$ mass sideband events and the cross-hatched
histogram shows the normalized continuum contributions.}
\label{GBdata}
\end{figure*}

\subsection{MEASUREMENTS OF $\chi_{b1} \to \omega+G_{0^{--}}$}

\begin{figure*}[htbp]
\includegraphics[height=6cm,angle=-90]{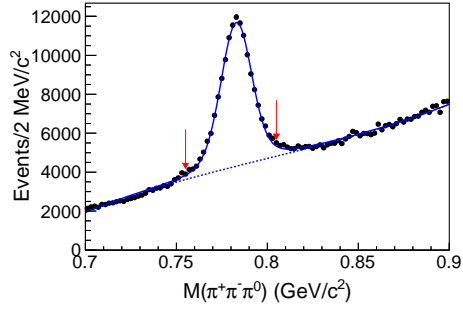}
\caption{The $\pi^+\pi^-\pi^0$ invariant mass distribution in $\Upsilon(2S)$ data. The arrows show
the $\omega$ signal region.} \label{omedata}
\end{figure*}

\begin{figure}[htpb]
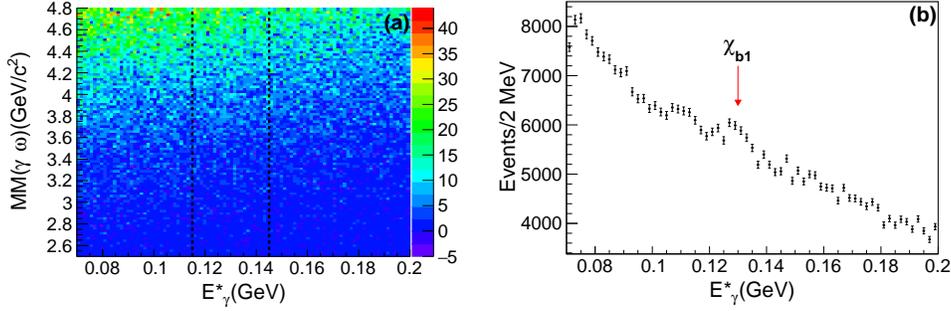

\includegraphics[height=6cm,angle=-90]{fig15a.epsi}
\hspace{0.35cm}
\includegraphics[height=6cm,angle=-90]{fig15b.epsi}
\caption{(Color online) Scatter plot of the recoil mass of
$\gamma \omega$ versus
the energy of the $\Upsilon(2S)$
radiative photon in the $\EE$ C.M. frame (a) and the distribution of the
energy of the $\Upsilon(2S)$ radiative photon (b). The dotted lines in (a) indicate the expected $\chi_{b1}$
signal region. The arrow in (b) shows the
position of $\chi_{b1}$.}\label{omegascatter}
\end{figure}

\begin{figure*}[htbp]
\includegraphics[height=7cm,angle=-90]{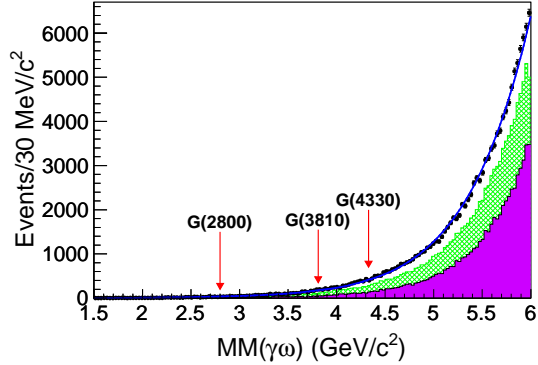}
\caption{(Color online)
The $\gamma\omega$ recoil mass spectrum for $\Upsilon(2S)\to \gamma \chi_{b1} \to \gamma \omega + anything$
in the $\Upsilon(2S)$ data sample.
The solid curve shows the result of the fit described in the text, including the
$G(2800)$, $G(3810)$, and $G(4330)$ states, with a common width fixed to 0.15 GeV and
with central values indicated by the arrows. The dashed curve shows the fitted background.
The shaded histogram is from the normalized $\omega$  mass sideband events and the cross-hatched
histogram shows the normalized continuum contributions.}
\label{omegarec}
\end{figure*}

Candidate $\omega$  states are reconstructed via $\pp\pi^0$.
We perform a
mass-constrained kinematic fit to the selected $\pi^0$ candidate
and require $\chi^{2}<10$. To remove the backgrounds with
$K^{0}_{S}$, the $\pi^+\pi^-$ invariant mass must not lie
between 0.475 and 0.515~GeV/$c^2$. As shown in Fig.~\ref{omedata}, a clear $\omega$ signal is seen
in the $\pi^+\pi^-\pi^0$ invariant mass spectrum in $\Upsilon(2S)$
data. We define the $\omega$ signal region as
$0.755~\hbox{GeV}/c^2<M(\pi^+\pi^-\pi^0)<0.805$~GeV/$c^2$ and its mass sideband
as $0.820~\hbox{GeV}/c^2<M(\pi^+\pi^-\pi^0)<0.870$~GeV/$c^2$.
Figure~\ref{omegascatter} shows the scatter plot of
the recoil mass of $\gamma \omega$ versus
the energy of
the $\Upsilon(2S)$ radiative photon in the $\EE$ C.M. and the distribution of the
energy of the $\Upsilon(2S)$ radiative photon. From the plots, no clear
$\chi_{b1}$ signal is observed. Figure~\ref{omegarec} shows the
recoil mass spectrum of $\gamma \omega$ for events in the $\omega$
signal region, and the background from the normalized $\omega$
mass sideband events and from the normalized continuum
contributions. No evident $G_{0^{--}}$ signal is observed. An
unbinned extended maximum-likelihood fit is applied to the $\gamma
\omega$ recoil mass spectrum.
The result of a fit including $G(2800)$,
$G(3810)$, and $G(4330)$ signals with widths fixed at
0.15~GeV is shown in Fig.~\ref{omegarec}. The fit yields $22.0\pm
34.1$ $G(2800)$, $129.6\pm 75.2$ $G(3810)$, and $132.9\pm 364.5$
$G(4330)$ signal events.

\begin{figure*}[htbp]
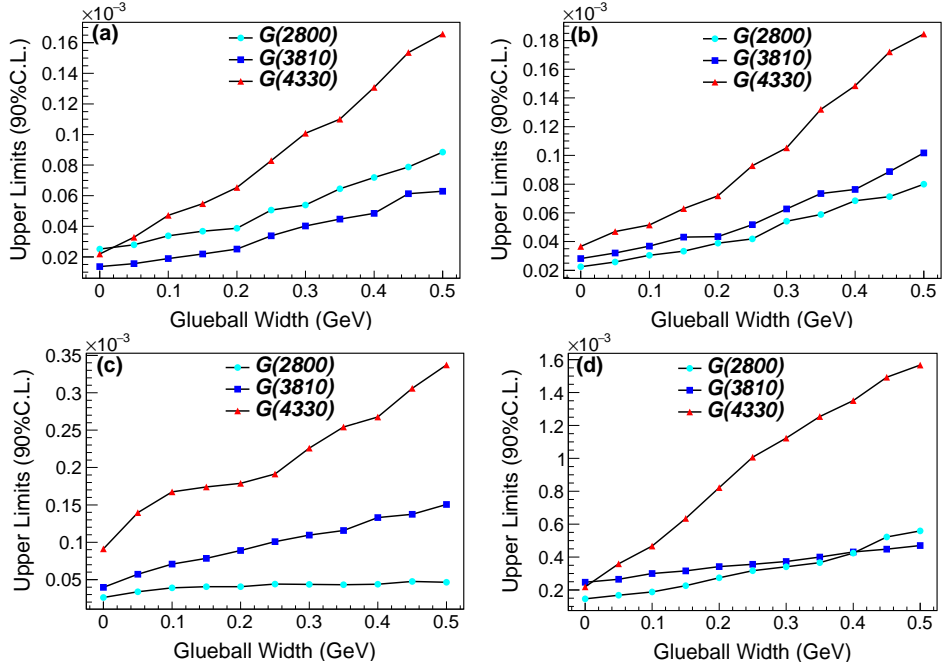

\includegraphics[height=6cm,angle=-90]{fig17a.epsi}
\hspace{0.3cm}
\includegraphics[height=6cm,angle=-90]{fig17b.epsi}
\hspace{0.3cm}
\includegraphics[height=6cm,angle=-90]{fig17c.epsi}
\hspace{0.3cm}
\includegraphics[height=5.8cm,width=4.38cm,angle=-90]{fig17d.epsi}
\caption{(Color online) The upper limits on the branching fractions for
$\Upsilon(1S)\to f_1(1285)+G_{0^{--}}$ (a), $\Upsilon(2S) \to
f_1(1285)+G_{0^{--}}$ (b), $\chi_{b1} \to \jpsi+G_{0^{--}}$ (c),
and $\chi_{b1} \to \omega+G_{0^{--}}$ (d)
as a function of the assumed $G_{0^{--}}$ decay width.}\label{final22}
\end{figure*}

Using the same method as described for $\Upsilon(1S,2S)\to
\chi_{c1}+G_{0^{--}}$, the calculated upper limits on the numbers
of signal events ($N^{\rm UL}$), the reconstruction efficiencies
($\varepsilon$), and the systematic uncertainties ($\sigma_{\rm
syst}$) for $\Upsilon(1S,2S)\to f_1(1285)+G_{0^{--}}$,
$\chi_{b1}\to J/\psi+G_{0^{--}}$ and $\chi_{b1}\to
\omega+G_{0^{--}}$ with different $G_{0^{--}}$ widths from 0.0 to
0.5~GeV in steps of 0.05~GeV are listed in
Table~\ref{Table-X-Summary}.
The results are displayed graphically in Fig.~\ref{final22}.


\section{SYSTEMATIC ERRORS}

Several sources of systematic errors are taken into account in the
branching fraction measurements.
The systematic uncertainty of 0.35\% per track
due to charged-track reconstruction is determined from a study
of partially reconstructed $D^{\ast+}\to D^0(\to K_S^0\pi^+\pi^-)\pi^+$ decays.
It is additive. The photon reconstruction contributes 2.0\% per photon, as determined using radiative Bhabha events. Based on the
measurements of the particle identification efficiencies of lepton
pairs from $\gamma\gamma\to \ell^+\ell^-$ events and pions from a
low-background sample of $D^\ast$ events, the MC simulation yields
uncertainties of $3.6\%$ for each lepton pair and $1.3\%$ for each
pion. The MC statistical errors are estimated using
the numbers of selected and generated
events;  these are $1.0\%$ or less. The trigger efficiency evaluated from simulation is
approximately 100\% with a negligible uncertainty. Errors on the branching fractions of the
intermediate states are taken from Ref.~\cite{PDG}. The uncertainties of the branching fractions of $\Upsilon(2S) \to \gamma \chi_{b1}$, $\chi_{c1} \to \gamma \jpsi$, $\jpsi \to \ell^+ \ell^-$, $f_1(1285) \to a_0(980) \pi$, $\eta \to \gamma \gamma$, $\omega \to \pi^+ \pi^- \pi^0$ and $\pi^0 \to \gamma\gamma$ are 5.8\%, 3.5\%, 1.1\%, 19.4\%, 0.5\%, 0.8\% and 0.04\%, respectively. By changing the order of the background polynomial
and the range of the fit, the decay-dependent relative difference in the upper
limits of the number of signal events is obtained; this
is taken as the systematic error due to the
uncertainty of the fit.
Finally, the uncertainties on the total numbers of $\Upsilon(1S)$
and $\Upsilon(2S)$ events are 2.2\% and 2.3\%, respectively, which
are mainly due to imperfect simulations of the charged-track
multiplicity distributions from inclusive hadronic MC events.
Assuming that all of these systematic-error sources are
independent, the total systematic errors are summed in quadrature
and listed in Table~\ref{Table-X-Summary} for all the studied
modes under the assumptions of different $G_{0^{--}}$ widths.

\section{Results and Discussion}

In summary, using the large data samples of $102\times 10^6$
$\Upsilon(1S)$ and $158\times 10^6$ $\Upsilon(2S)$ events
collected by the Belle detector, we have searched for the $0^{--}$ glueball
in $\Upsilon(1S)$, $\Upsilon(2S)$, and $\chi_{b1}$ decays for the
first time. No evident signal is found at three theoretically-predicted
masses in the processes $\Upsilon(1S,2S)\to
\chi_{c1}+G_{0^{--}}$, $\Upsilon(1S,2S)\to f_1(1285)+G_{0^{--}}$,
$\chi_{b1}\to \jpsi+G_{0^{--}}$,  and $\chi_{b1}\to
\omega+G_{0^{--}}$ and $90\%$ C.L. upper limits are set on the
branching fractions for these processes. Figures~\ref{final1}
and~\ref{final22} show the upper limits
on the branching fractions as a function of the $0^{--}$ glueball
width. The results presented in this article do not strongly depend on the spin-parity assumption of
the glueballs.
We also scan with fits across the mass regions up to 6.0 GeV/$c^2$ for
all of the modes under study. All the signal significances are less than 3$\sigma$ except for $\Upsilon(1S) \to f_1(1285)+G_{0^{--}}$,
where the maximum signal significance is 3.7$\sigma$ at 3.92~GeV/$c^2$.
It should be noted that we report here the local statistical significances without
considering the look-elsewhere effect, which will largely reduce the significances.
As we do not observe signals in any of the modes under study,
the upper limits can be applied almost directly to the glueballs in this mass region with
the same width and opposite spin parity and charge-conjugate parity, such as
$J^{PC}=(0,1,2,3)^{+-}$ and $(1,2,3)^{--}$~\cite{Chen:2005mg}.
In addition, distinct $\chi_{c1}$ signals are observed in
the $\Upsilon(1S)$ and $\Upsilon(2S)$ inclusive decays. The
corresponding branching fractions are measured to be
$\BR(\Upsilon(1S)\to \chi_{c1}+ anything) =
(1.90\pm 0.43(stat.)\pm 0.14(syst.))\times 10^{-4}$ with substantially
improved precision compared to the previous result of $(2.3\pm
0.7)\times 10^{-4}$~\cite{PhysRevD.70.072001}, and
$\BR(\Upsilon(2S)\to \chi_{c1}+ anything) =
(2.24\pm 0.44(stat.)\pm 0.20(syst.))\times 10^{-4}$, measured for the first time.

\begin{table*}[t]
\caption{\label{Table-X-Summary} Summary of the upper limits for
$\Upsilon(1S,2S)\to \chi_{c1}+G_{0^{--}}$, $f_1(1285)+G_{0^{--}}$,
and $\chi_{b1} \to J/\psi+G_{0^{--}}$, $\omega+G_{0^{--}}$ under
different assumptions of $G_{0^{--}}$ width ($\Gamma$ in GeV),
where $N^{\rm UL}$ is the upper limit on the number of signal
events taking into account systematic errors, $\varepsilon$ is the
reconstruction efficiency, $\sigma_{\rm syst}$ is the total
systematic uncertainty and $\BR^{\rm UL}$ is the 90\% C.L. upper limit on the branching
fraction.}
\scriptsize
  \begin{tabular}{c|cccc|cccc}
  \hline\hline
    \multicolumn{5}{c}{$\Upsilon(1S) \to \chi_{c1}+G(2800)/G(3810)/G(4330)$} & \multicolumn{4}{c}{$\Upsilon(2S) \to \chi_{c1}+G(2800)/G(3810)/G(4330)$}\\
    $\Gamma$  & $N^{\rm UL}$ & $\varepsilon (\%)$ & $\sigma_{\rm syst} (\%)$ & $\BR^{\rm UL}(\times 10^{-6})$ & $N^{\rm UL}$ & $\varepsilon (\%)$ & $\sigma_{\rm syst}(\%)$ & $\BR^{\rm UL}(\times 10^{-6})$ \\
  \hline
    $0.00$& 5.5/4.4/9.2  & 19.9/24.6/26.3 &   6.6/10.8/7.6 & 6.8/4.5/8.8       &12.9/6.9/11.3&  19.7/24.6/25.6&   6.6/15.9/8.6  &10.6/4.4/7.3  \\
    $0.05$& 6.1/5.6/11.1  & 19.5/25.1/26.5&  6.7/8.1/9.6  &7.8/5.3/10.6          &14.7/8.2/16.6&  19.2/24.7/25.6&   6.7/14.2/16.4 &12.7/5.5/10.8  \\
    $0.10$& 6.8/7.0/13.3  & 20.2/24.6/26.0&  7.1/7.2/11.4  &8.6/7.2/13.0        &16.2/9.5/23.7&  19.8/24.5/25.2& 7.0/16.8/21.5  &13.5/6.4/15.6 \\
    $0.15$& 7.3/9.9/15.2  &  20.1/25.0/26.4&  7.3/6.3/12.9 &9.1/10.0/14.6       &16.9/10.9/30.6&  19.6/24.5/26.0& 7.4/18.9/21.8 &14.4/7.3/19.6  \\
    $0.20$&  7.6/11.6/17.2  &  19.8/25.0/25.9&  7.3/6.3/13.9 &9.8/11.8/16.8         &17.0/11.8/38.4&  19.7/24.0/25.2 &7.6/20.6/27.1&14.4/8.0/25.8   \\
    $0.25$&8.5/14.5/21.4  &  19.6/24.5/26.4&7.4/6.3/16.5 &10.9/15.1/20.6          &18.7/14.6/47.1&  18.8/24.0/25.6 &   8.6/23.8/28.4 &16.6/10.0/30.6   \\
    $0.30$&8.7/16.3/24.9   &  18.9/24.4/26.6& 6.7/6.5/16.2&11.6/17.0/23.8         &20.6/16.5/54.9& 18.6/23.8/25.7  &  11.4/26.4/29.7 &18.4/11.5/34.9  \\
    $0.35$&  8.9/19.3/28.2  &  19.6/24.3/26.6&  6.3/7.2/19.8&11.8/20.0/26.8       &20.1/18.0/62.5& 18.8/23.7/25.6  &  12.5/27.3/34.2&17.8/12.5/39.9  \\
    $0.40$&  9.0/21.8/29.7  &  19.2/24.8/26.3&  10.3/7.3/20.3&12.0/22.3/28.7      &21.4/21.3/75.7& 18.6/24.2/25.8  &   9.2/31.0/42.3&19.0/14.6/48.9 \\
    $0.45$&  9.2/22.7/32.3  &  19.2/24.3/26.4&  10.1/7.5/21.2&12.2/23.7/31.1      &21.5/22.2/85.6& 18.8/23.9/25.2  &  9.9/31.7/44.4&19.1/15.5/56.5 \\
    $0.50$&  9.6/24.2/36.8  &  19.4/24.8/26.8 &  7.7/8.1/22.7&12.5/24.7/35.0      &22.4/24.1/103.7& 18.7/24.0/25.9  & 11.7/33.0/47.8&19.9/16.6/66.8    \\
  \hline\hline

  \multicolumn{5}{c}{$\Upsilon(1S) \to f_1(1285)+G(2800)/G(3810)/G(4330)$} & \multicolumn{4}{c}{$\Upsilon(2S) \to f_1(1285)+G(2800)/G(3810)/G(4330)$} \\
    $\Gamma$  & $N^{\rm UL}$ & $\varepsilon (\%)$ & $\sigma_{\rm syst} (\%)$ & $\BR^{\rm UL}(\times 10^{-5})$ & $N^{\rm UL}$ & $\varepsilon (\%)$ & $\sigma_{\rm syst}(\%)$ & $\BR^{\rm UL}(\times 10^{-5})$ \\
  \hline/
    $0.00$& 23.0/19.5/33.0  & 8.3/9.9/10.4 &22.5/22.4/23.0&2.5/1.4/2.2   &38.6/61.1/83.4&  7.7/9.7/10.2&20.5/22.4/22.1 & 2.2/2.8/3.7  \\
    $0.05$& 33.4/22.4/49.7  & 8.3/9.9/10.5&22.7/21.0/25.2&2.8/1.6/3.3    &45.1/69.8/107.3& 7.8/9.7/10.2&20.7/20.4/23.2 & 2.6/3.2/4.7   \\
    $0.10$& 40.3/26.9/70.5  & 8.2/9.8/10.3&23.3/24.7/28.3&3.4/1.9/4.7    &53.6/80.1/118.6&  7.8/9.7/10.2&20.9/21.1/25.0& 3.1/3.7/5.2  \\
    $0.15$& 43.0/31.6/83.0  & 8.1/10.0/10.5&24.0/31.5/29.5&3.7/2.2/5.5   &58.6/92.4/143.2& 7.8/9.6/10.1&21.0/21.4/24.3 & 3.3/4.3/6.3  \\
    $0.20$& 45.7/35.8/97.2  & 8.2 /9.9/10.3&24.4/33.6/32.5&3.9/2.5/6.5     &68.2/92.8/165.5&  7.8/9.5/10.3 &21.2/22.0/24.6 & 3.9/4.3/7.2 \\
    $0.25$&59.8/48.0/123.6  & 8.2/9.8/10.3&26.4/27.5/34.4&5.1/3.4/8.8      &73.4/110.7/213.3&  7.8/9.5/10.2 &21.4/22.6/25.3& 4.2/5.2/9.3 \\
    $0.30$&63.4/57.1/152.3   &  8.1/9.8/10.4&26.8/25.4/35.9&5.4/4.0/10.0    &95.0/134.2/239.4& 7.8/9.6/10.1  &21.9/21.7/25.7& 5.4/6.3/10.5  \\
    $0.35$& 74.8/63.3/163.7&  8.0/9.8/10.3& 27.7/22.5/36.8&6.5/4.5/11.0     &101.3/156.9/299.2& 7.7/9.5/10.1  &22.1/21.2/25.6& 5.9/7.3/13.2 \\
    $0.40$& 82.1/68.3/195.1&  7.9/9.7/10.3&29.3/22.2/36.8&7.2/4.9/13.1      &119.6/165.8/337.5& 7.8/9.7/10.1  &22.7/20.4/25.6& 6.8/7.6/14.8 \\
    $0.45$& 90.4/86.5/229.4&  7.9/9.7/10.3&30.2/20.3/38.5&7.9/6.1/15.4      &120.4/187.4/388.4& 7.5/9.4/10.1  &22.7/23.2/26.0& 7.1/8.9/17.2 \\
    $0.50$& 103.8/89.1/248.1&  8.1/9.8/10.3 &30.4/23.0/38.7&8.8/6.3/16.6    &135.8/214.6/416.3& 7.6/9.4/10.1  &23.3/22.5/26.0& 8.0/10.2/18.4  \\
  \hline\hline

    \multicolumn{5}{c}{$\chi_{b1} \to \jpsi+G(2800)/G(3810)/G(4330)$}& \multicolumn{4}{c}{$\chi_{b1} \to \omega+G(2800)/G(3810)/G(4330)$} \\
    $\Gamma$  & $N^{\rm UL}$ & $\varepsilon (\%)$ & $\sigma_{\rm syst} (\%)$ & $\BR^{\rm UL}(\times 10^{-5})$ & $N^{\rm UL}$ & $\varepsilon (\%)$ & $\sigma_{\rm syst}(\%)$ & $\BR^{\rm UL}(\times 10^{-4})$ \\
  \hline
    $0.00$&5.9/11.4/29.4& 17.8/23.6/26.2 &9.4/9.5/21.2 &2.6/4.0/9.1  &57.7/132.7/133.5& 4.1/5.6/6.3 &9.8/11.0/14.4&1.4/2.5/2.2   \\
    $0.05$&7.8/15.8/43.6& 18.3/22.7/25.6 &9.6/9.9/15.3 &3.4/5.7/13.9  &66.2/148.2/223.7& 4.1/5.8/6.5&9.7/10.3/9.4 &1.7/2.6/3.6   \\
    $0.10$&8.9/19.6/51.4& 18.4/22.6/25.0 &9.2/10.0/14.6&3.9/7.1/16.7  &74.0/161.4/285.9& 4.1/5.6/6.4 &10.3/14.9/9.0&1.8/3.0/4.7    \\
    $0.15$&9.3/22.3/55.6& 18.2/23.1/26.2 &9.2/8.3/13.6 &4.0/7.9/17.4  &91.1/166.6/384.5& 4.2/5.5/6.3&9.8/19.1/8.5 &2.2/3.2/6.3  \\
    $0.20$&9.5/25.5/56.6& 18.5/23.4/25.9 &9.2/7.8/13.0 &4.1/8.9/17.8  &110.0/178.6/494.9& 4.2/5.4/6.2&9.7/20.2/12.3&2.7/3.4/8.2   \\
    $0.25$&9.7/29.8/60.8& 18.0/24.1/25.9 &8.3/7.8/12.7 &4.2/10.1/19.1  &119.5/185.7/603.9& 4.0/5.4/6.2&9.4/20.9/9.3&3.2/3.6/10.1     \\
    $0.30$&9.8/31.9/71.5& 17.8/23.8/25.9 &9.3/8.0/13.1 &4.3/10.9/22.6  &131.9/200.3/686.3& 4.0/5.6/6.4&9.9/22.0/10.8&3.4/3.7/11.2    \\
    $0.35$&9.9/34.0/77.9&18.2/24.0/25.1 &10.8/8.1/14.0 &4.4/11.6/25.4  &144.6/210.8/761.1&4.1/5.5/6.3&10.8/19.7/8.9&3.6/4.0/12.5   \\
    $0.40$&9.9/38.6/83.5&18.0/23.7/25.5 &9.2/8.5/13.4  &4.5/13.3/26.7  &164.1/226.4/814.8&4.0/5.4/6.3&11.6/17.2/17.7& 4.2/4.3/13.5    \\
    $0.45$&10.3/38.9/95.5& 17.8/23.2/25.6 &8.9/8.5/13.6&4.6/13.7/30.6  &201.9/235.6/906.0& 4.0/5.4/6.3 &11.8/16.1/10.4&5.2/4.5/14.9  \\
    $0.50$&10.4/42.9/105.1&18.1/23.3/25.5 &10.1/8.5/14.0&4.7/15.0/33.7  &209.0/244.4/983.7&3.9/5.4/6.5 &11.6/11.7/9.3 &5.6/4.7/15.7    \\
  \hline\hline
  \end{tabular}
\end{table*}

\section{ACKNOWLEDGMENTS}
We thank the KEKB group for the excellent operation of the
accelerator; the KEK cryogenics group for the efficient
operation of the solenoid; and the KEK computer group,
the National Institute of Informatics, and the
PNNL/EMSL computing group for valuable computing
and SINET5 network support.  We acknowledge support from
the Ministry of Education, Culture, Sports, Science, and
Technology (MEXT) of Japan, the Japan Society for the
Promotion of Science (JSPS), and the Tau-Lepton Physics
Research Center of Nagoya University;
the Australian Research Council;
Austrian Science Fund under Grant No.~P 26794-N20;
the National Natural Science Foundation of China under Contracts
No.~10575109, No.~10775142, No.~10875115, No.~11175187, No.~11475187,
No.~11521505 and No.~11575017;
the Chinese Academy of Science Center for Excellence in Particle Physics;
the Ministry of Education, Youth and Sports of the Czech
Republic under Contract No.~LG14034;
the Carl Zeiss Foundation, the Deutsche Forschungsgemeinschaft, the
Excellence Cluster Universe, and the VolkswagenStiftung;
the Department of Science and Technology of India;
the Istituto Nazionale di Fisica Nucleare of Italy;
the WCU program of the Ministry of Education, National Research Foundation (NRF)
of Korea Grants No.~2011-0029457,  No.~2012-0008143,
No.~2014R1A2A2A01005286,
No.~2014R1A2A2A01002734, No.~2015R1A2A2A01003280,
No.~2015H1A2A1033649, No.~2016R1D1A1B01010135, No.~2016K1A3A7A09005603, No.~2016K1A3A7A09005604, No.~2016R1D1A1B02012900,
No.~2016K1A3A7A09005606, No.~NRF-2013K1A3A7A06056592;
the Brain Korea 21-Plus program and Radiation Science Research Institute;
the Polish Ministry of Science and Higher Education and
the National Science Center;
the Ministry of Education and Science of the Russian Federation and
the Russian Foundation for Basic Research;
the Slovenian Research Agency;
Ikerbasque, Basque Foundation for Science and
the Euskal Herriko Unibertsitatea (UPV/EHU) under program UFI 11/55 (Spain);
the Swiss National Science Foundation;
the Ministry of Education and the Ministry of Science and Technology of Taiwan;
and the U.S.\ Department of Energy and the National Science Foundation.



\begin{thebibliography}{**}
\bibitem{QCD} R. L. Jaffe and K. Johnson, Phys. Lett. B {\bf 60}, 201 (1976).
\bibitem{5} C. F. Qiao and L. Tang, Phys. Rev. Lett. {\bf 113}, 221601 (2014).
\bibitem{1507} L. Bellantuono, P. Colangelo and F. Giannuzzi, J. High Energy Phys. {\bf 1510}, 137 (2015).
\bibitem{number} C. P. Shen {\em et al.} (Belle Collaboration), Phys. Rev. D {\bf 88}, 011102(R) (2013).
\bibitem{Abashian2002117} A.~Abashian {\em et al.} (Belle Collaboration),
 Nucl. Instr. and Methods Phys. Res. Sect. A {\bf 479}, 117 (2002).
\bibitem{PTEP201204D001} J. Brodzicka {\em et al.}, Prog. Theor. Exp. Phys. (2012) 04D001.
\bibitem{Kurokawa20031} S. Kurokawa and E. Kikutani, Nucl. Instr. and Methods Phys. Res. Sect. A {\bf 499}, 1 (2003), and other papers
included in this volume.
\bibitem{PTEP201303A001} T. Abe {\em et al.}, Prog. Theor. Exp. Phys. (2013) 03A001 and following articles up to 03A011.
\bibitem{Lange2001152} D. J. Lange, Nucl. Instr. and Methods Phys. Res. Sect. A {\bf 462}, 152 (2001).
\bibitem{mcang} K. W. Edwards  {\em et al.} (CLEO Collaboration), Phys. Rev. D {\bf 59}, 032003 (1999).
\bibitem{JHEP2006.026} T. Sjostrand, S. Mrenna and P. Skands, J. High Energy Phys. {\bf 05}, 026 (2006).
\bibitem{like} E. Nakano, Nucl. Instr. and Methods Phys. Res. Sect. A {\bf 494}, 402 (2002).
\bibitem{Hanagaki485490} K. Hanagaki {\em et al.},  Nucl. Instr. and Methods Phys. Res. Sect. A {\bf 485}, 490 (2002).
\bibitem{Abashian49169} A. Abashian {\it et al.}, Nucl. Instr. and Methods Phys. Res. Sect. A {\bf 491}, 69 (2002).
\bibitem{jpsimass} V.~M.~Aulchenko {\em et al.}, (KEDR Collaboration), Phys. Rett. B {\bf 573}, 63 (2003).
\bibitem{PhysRevD.69.094027} K. Y. Liu, Z. G. He and K. T. Chao, Phys. Rev. D {\bf 69}, 094027 (2004).
\bibitem{PhysRevD.56.321} F. Yuan, C. F. Qiao and K. T. Chao, Phys. Rev. D {\bf 56}, 321 (1997).
\bibitem{hep-hp13108597}  K. T. Chao, Z. G. He, D. Li and C. Meng, arXiv:1310.8597 [hep-ph]; C. Meng (private communication).
\bibitem{PhysRevD.70.072001}  R. A. Briere  {\it et al.} (CLEO Collaboration), Phys. Rev. D {\bf 70}, 072001 (2004).
\bibitem{PDG} K. A.~Olive  {\it et al.} (Particle Data Group), Chin. Phys. C {\bf 38}, 090001 (2014).
\bibitem{Nov} The Novosibirsk function is defined as $f(x) =
\exp[-\frac12(\ln^2(1+\Lambda(x-x_0))/\tau^2+\tau^2)]$ with
$\Lambda=$ $\sinh(\tau\sqrt{\ln 4})/(\sigma\sqrt{\ln4})$. The
parameters represent the mean ($x_0$), the width ($\sigma$) and
the tail asymmetry ($\tau$).
\bibitem{cl} In common high-energy physics usage, this
Bayesian interval has been reported as ``confidence interval,'' which
is a frequentist-statistics term.
\bibitem{Chen:2005mg} Y.~Chen {\it et al.}, Phys.\ Rev.\ D {\bf 73}, 014516 (2006).


\end{thebibliography}
\end{document}